\documentstyle[preprint,aps,epsfig,rotate]{revtex}
\input epsf

\def\ETslash{\not{\hbox{\kern-4pt $E_T$}}}

\def\ra{\rightarrow}

\def\be{\begin{equation}}
\def\ee{\end{equation}}
\def\bea{\begin{eqnarray}}
\def\eea{\end{eqnarray}}

\def\D0{D\O~}

\def\ra{\rightarrow}

\newfont{\titlefont}{cmbx10 scaled 1270}      
\newfont{\titlefonta}{cmbx10 scaled 1100}      
\catcode`\@=11
\long\def\@makefntext#1{
\protect\noindent \hbox to 3.2pt {\hskip-.9pt
$^{{\ninerm\@thefnmark}}$\hfil}#1\hfill}                
\def\@makefnmark{\hbox to 0pt{$^{\@thefnmark}$\hss}}  
\def\ps@myheadings{\let\@mkboth\@gobbletwo
\def\@oddhead{\hbox{}
\rightmark\hfil\ninerm\thepage}
\def\@oddfoot{}\def\@evenhead{\ninerm\thepage\hfil
\leftmark\hbox{}}\def\@evenfoot{}
\def\sectionmark##1{}\def\subsectionmark##1{}}
\renewcommand{\thefootnote}{\fnsymbol{footnote}}

\renewenvironment{thebibliography}[1]
    {\begin{list}{\arabic{enumi}.}
    {\usecounter{enumi}\setlength{\parsep}{0pt}
\setlength{\leftmargin 1.25cm}{\rightmargin 0pt}
     \setlength{\itemsep}{0pt} \settowidth
    {\labelwidth}{#1.}\sloppy}}{\end{list}}
\newcounter{itemlistc}
\newcounter{romanlistc}
\newcounter{alphlistc}
\newcounter{arabiclistc}

\newcommand{\tcaption}[1]{
        \refstepcounter{table}
        \setbox\@tempboxa = \hbox{\footnotesize Table~\thetable. #1}
        \ifdim \wd\@tempboxa > 6in
           {\begin{center}
        \parbox{6in}{\footnotesize\baselineskip=15pt Table~\thetable. #1}
            \end{center}}
        \else
             {\begin{center}
             {\footnotesize Table~\thetable. #1}
              \end{center}}
        \fi}

\newcommand{\fcaption}[1]{
        \refstepcounter{figure}
        \setbox\@tempboxa = \hbox{\footnotesize Fig.~\thefigure. #1}
        \ifdim \wd\@tempboxa > 6in
           {\begin{center}
        \parbox{6in}{\footnotesize\baselineskip=12pt Fig.~\thefigure. #1}
            \end{center}}
        \else
             {\begin{center}
             {\footnotesize Fig.~\thefigure. #1}
              \end{center}}
        \fi}

 1 
 1  
 1  
 1  
 1  
 1  
 1  

\font\ninerm=cmr9

\evensidemargin=\oddsidemargin  
\renewcommand{\baselinestretch}{1}
\def\today{\number\day
           \space\ifcase\month\or
             January\or February\or March\or April\or May\or June\or
             July\or August\or September\or October\or November\or December\fi
           \space\number\year}

\evensidemargin=\oddsidemargin  
\def\doublespaced{\baselineskip=\normalbaselineskip\multiply
    \baselineskip by 150\divide\baselineskip by 100}
\doublespaced

\begin{document}

\addtolength{\textheight}{2.4cm}
\addtolength{\topmargin}{1.8cm}
\evensidemargin=\oddsidemargin
\renewcommand{\thefootnote}{\fnsymbol{footnote}}
\setcounter{footnote}{0}
\begin{flushright}
TUIMP-TH-99/106
\end{flushright}

\null\vspace{0.5cm}

\centerline{{\large\bf Physics Overview}\footnote{Talk presented at {\it
The First ACFA Workshop on Physics/Detector at the Linear Collider},
Nov. 26-27, 1998, Tsinghua Univ. Beijing, China.}}


\vspace*{0.4cm}
\centerline{ {\sc Yu-Ping Kuang} }
  
\vspace*{0.1cm}

\baselineskip=17pt

\centerline{\it
Department of Physics, Tsinghua University, Beijing 100084, China}

\renewcommand{\baselinestretch}{1}
\begin{abstract} 
\noindent
Recent developments of physics at the TeV energy scale, especially physics 
related to the $e^+e^-$ linear colliders are briefly reviewed. The
topics include the present status of the standard model, Higgs physics, 
supersymmetry, strongly interacting electroweak symmetry breaking mechanism, 
and top quark physics.
\\
\end{abstract}


\addtolength{\textheight}{-2.4cm}
\addtolength{\topmargin}{0.0cm}
\addtolength{\textheight}{0.5cm}

\normalsize
\baselineskip=19.5pt
\setcounter{footnote}{00}
\renewcommand{\thefootnote}{\arabic{footnote}}
\renewcommand{\baselinestretch}{1}

\addtolength{\topmargin}{0.2cm}
\addtolength{\textheight}{-0.4cm}
\baselineskip=19.5pt
\renewcommand{\baselinestretch}{1}

\null\vspace{0.1cm}
\null\noindent
\begin{center}
{\bf I. Introduction}
\end{center}

\vspace{0.2cm}

One of the most remarkable results in recent particle physics is that
the Standard Model (SM) has successfully passed the recent precision 
experimental tests at the CERN LEP, the SLAC SLC and the Fermilab
Tevatron with the LEP and SLC tests of the precision of one-loop 
electroweak radiative corrections \cite{EL,Hollik,SLD,top}.
Theoretically, this means that the present available tests support the 
renormalizable $SU(2)\times U(1)$ gauge theory of the SM as the 
theory of electroweak interactions. An important ingredient of such a
theory is the spontaneous breakdown of the $SU(2)\times U(1)$ gauge symmetry.

Despite the present success of the SM, the mechanism of the electroweak 
symmetry breaking (EWSB) is not clearly known yet. In the
$SU(2)\times U(1)$ electroweak theory, all particle masses come from
the EWSB sector. Thus probing the EWSB mechanism concerns the
understanding of the {\it origin of all particle masses}, which is a
very deep and fundamental problem in physics. In the SM, it is assumed 
that an elementary Higgs field is responsible for the EWSB. After careful 
experimental searches at LEP, the SM Higgs boson has 
not been found, and the present data are not so sensitive to the Higgs mass
$m_H$ due to Veltman's screening theorem~\cite{screening}. 
From the theoretical point of view, there are several unsatisfactory features
in the Higgs sector of the SM, e.g. there are so many free parameters related
to the Higgs sector, and there are the well-known problems of {\it triviality}
and {\it unnaturalness} \cite{Chanowitz,Susskind}. Usually, people take the 
point of view that the present theory of the SM is only valid up to a certain 
energy scale $\Lambda$, and new physics will become important above 
$\Lambda$. Possible new physics are supersymmetry (SUSY) and dynamical
EWSB mechanism concerning new strong interactions. So that probing the 
mechanism of EWSB also concerns the discovery of new physics beyond the SM. 
Such an important problem can be experimentally studied at LEP and the future 
TeV energy colliders such as the upgraded Fermilab Tevatron, the CERN LHC, and 
the future $e^+e^-$ liner colliders.

In this paper, we give a short theoretical review on the recent developments 
of physics at the TeV energy scale, especially on physics related to the
$e^+e^-$ linear colliders. We shall first briefly review the present status
of the SM in Sec.II as the basis of the other parts of this review. In
Sec. III, we shall deal with the physics of the Higgs boson. Sec. IV is a
brief review on SUSY, and Sec. V concerns some
recent developments of the study of strongly interacting EWSB mechanisms.
In Sec. VI, we shall discuss some topics related to the top quark. The
conclusions will be given in Sec. VII. 

\null\vspace{1cm}
\begin{center}
{\bf II. Status of the Standard Model}
\end{center}

Recent data from LEP, SLC and the Tevatron supports the electroweak SM
as the theory of electroweak interactions to the precision of one-loop
radiative corrections. There are recent reports on this subject 
\cite{EL,Hollik,SLD}. Here we quote a table of the global fit results
of the $Z$ pole precision observables given in Ref.\cite{EL} to show the 
situation (Table I).

\tcaption{$Z$ pole precision observables from LEP and the SLC.
Shown are the experimental results, the SM predictions, and the pulls.
The SM errors are from the uncertainties in $M_Z$, $\ln M_H$, $m_t$, 
$\alpha (M_Z)$, and $\alpha_s$. They have been treated as Gaussian 
and their correlations have been taken into account. Quoted from 
Ref.\cite{EL}.} 
\label{zpole}
\vspace{0.2cm}
\begin{center}
\footnotesize
\begin{tabular}{|l|c|c|c|r|}
\hline Quantity & Group(s) & Value & Standard Model & pull \\ 
\hline
$M_Z$ \hspace{14pt}      [GeV]&     LEP     &$ 91.1867 \pm 0.0021 $&$ 91.1865 \pm 0.0021 $&$ 0.1$ \\
$\Gamma_Z$ \hspace{17pt} [GeV]&     LEP     &$  2.4939 \pm 0.0024 $&$  2.4957 \pm 0.0017 $&$-0.8$ \\
$\Gamma({\rm had})$\hspace{8pt}[GeV]&  LEP  &$  1.7423 \pm 0.0023 $&$  1.7424 \pm 0.0016 $&  ---  \\
$\Gamma({\rm inv})$\hspace{11pt}[MeV]& LEP  &$500.1    \pm 1.9    $&$501.6    \pm 0.2    $&  ---  \\
$\Gamma({\ell^+\ell^-})$ [MeV]&     LEP     &$ 83.90   \pm 0.10   $&$ 83.98   \pm 0.03   $&  ---  \\
$\sigma_{\rm had}$ \hspace{12pt}      [nb] &     LEP     &$ 41.491  \pm 0.058  $&$ 41.473  \pm 0.015  $&$ 0.3$ \\
$R_e$                         &     LEP     &$ 20.783  \pm 0.052  $&$ 20.748  \pm 0.019  $&$ 0.7$ \\
$R_\mu$                       &     LEP     &$ 20.789  \pm 0.034  $&$ 20.749  \pm 0.019  $&$ 1.2$ \\
$R_\tau$                      &     LEP     &$ 20.764  \pm 0.045  $&$ 20.794  \pm 0.019  $&$-0.7$ \\
$A_{FB} (e)$                  &     LEP     &$  0.0153 \pm 0.0025 $&$  0.0161 \pm 0.0003 $&$-0.3$ \\
$A_{FB} (\mu)$                &     LEP     &$  0.0164 \pm 0.0013 $&$                    $&$ 0.2$ \\
$A_{FB} (\tau)$               &     LEP     &$  0.0183 \pm 0.0017 $&$                    $&$ 1.3$ \\
\hline
$R_b$                         &  LEP + SLD  &$  0.21656\pm 0.00074$&$  0.2158 \pm 0.0002 $&$ 1.0$ \\
$R_c$                         &  LEP + SLD  &$  0.1735 \pm 0.0044 $&$  0.1723 \pm 0.0001 $&$ 0.3$ \\
$R_{s,d}/R_{(d+u+s)}$         &     OPAL    &$  0.371  \pm 0.023  $&$  0.3592 \pm 0.0001 $&$ 0.5$ \\
$A_{FB} (b)$                  &     LEP     &$  0.0990 \pm 0.0021 $&$  0.1028 \pm 0.0010 $&$-1.8$ \\
$A_{FB} (c)$                  &     LEP     &$  0.0709 \pm 0.0044 $&$  0.0734 \pm 0.0008 $&$-0.6$ \\
$A_{FB} (s)$                  &DELPHI + OPAL&$  0.101  \pm 0.015  $&$  0.1029 \pm 0.0010 $&$-0.1$ \\
$A_b$                         &     SLD     &$  0.867  \pm 0.035  $&$  0.9347 \pm 0.0001 $&$-1.9$ \\
$A_c$                         &     SLD     &$  0.647  \pm 0.040  $&$  0.6676 \pm 0.0006 $&$-0.5$ \\
$A_s$                         &     SLD     &$  0.82   \pm 0.12   $&$  0.9356 \pm 0.0001 $&$-1.0$ \\
\hline
$A_{LR}$ (hadrons)            &     SLD     &$  0.1510 \pm 0.0025 $&$  0.1466 \pm 0.0015 $&$ 1.8$ \\
$A_{LR}$ (leptons)            &     SLD     &$  0.1504 \pm 0.0072 $&$                    $&$ 0.5$ \\
$A_\mu$                       &     SLD     &$  0.120  \pm 0.019  $&$                    $&$-1.4$ \\
$A_\tau$                      &     SLD     &$  0.142  \pm 0.019  $&$                    $&$-0.2$ \\
$A_e (Q_{LR})$                &     SLD     &$  0.162  \pm 0.043  $&$                    $&$ 0.4$ \\
$A_\tau ({\cal P}_\tau)$      &     LEP     &$  0.1431 \pm 0.0045 $&$                    $&$-0.8$ \\
$A_e ({\cal P}_\tau)$         &     LEP     &$  0.1479 \pm 0.0051 $&$                    $&$ 0.3$ \\
$\bar{s}_\ell^2 (Q_{FB})$     &     LEP     &$  0.2321 \pm 0.0010 $&$  0.2316 \pm 0.0002 $&$ 0.5$ \\
\hline
\end{tabular}
\end{center}
\null\noindent
We see that the agreement of the SM with the recent data is quite good.
The deviations are all within $2\sigma$. In the SM fit analysis, the
top quark mass $m_t$ is taken as one of the fitting parameters. The best
fit requires $m_t=171.4\pm 4.8$ GeV \cite{EL} which is very close to the 
directly measured value at the Tevatron, $m_t=173.8\pm 3.2 (stat.)\pm 3.9
(syst.)$ GeV \cite{top,PDG}. This is a remarkable success of the SM. Such a 
good agreement supports the $SU(2)\times U(1)$ gauge interactions in the SM.

Despite of the above success of the SM, its EWSB sector is still not clear.
The assumed Higgs boson in the EWSB sector of the SM has not been found
yet. The experimental lower bound of the Higgs mass from the recent LEP 
experiments is $~m_H>94.1~$GeV \cite{DELPHI}.
Although the global fit analysis favors the SM with a light Higgs boson 
\cite{EL,Hollik}, there exist examples of models with new physics without 
a light Higgs boson, which can also be consistent with the present precision 
data \cite{strong,topseesaw}. So that whether there is a light Higgs
boson or not should be tested by future experimants, especially at the LHC and 
the LC.

Furhtermore, as has been pointed out in Refs.\cite{Field,Renton} that the
global fit includes a large number of 'raw' observables with large
errors, which may dilute some deviations of more precise observables.
In Refs.\cite{Field,Renton}, a sharper test of the SM is presented from
a model-independent analyses of the forward-backward asymmetry data in $Z$
decays. Instead of assuming the SM, they extract the {\it left-handed} and 
{\it right-handed} effective coupling constants of leptons and quarks from
the forward-backward asymmetry data by imposing a {\it weaker} assumption of
{\it lepton-quark universality}. The obtained results show that the SM fits
the extracted left-handed and right-handed couplings of the leptons,
and the $u,~d$ and $c$ quarks very well. However, for the $b$ quark, the 
extracted left-handed and right-handed couplings are \cite{Field}
\begin{eqnarray}
\bar{g}^L_b=-0.4159(24)\,,~~~~~~~~~~~~~~~~\bar{g}^R_b=0.1050(90)\,.
\end{eqnarray}
These are to be compared with the corresponding SM predictions up to
two-loop corrections: $\bar{g}^L_b=-0.4208$ and $\bar{g}^R_b=0.0774$. We
see that the agreement of $\bar{g}^L_b$ is good, while {\it the SM predicted 
$\bar{g}^R_b$ is smaller than the experimental value by 3.1$\sigma$} which is 
more significant than what we have seen in the global fit.

A similar deviation has also been mentioned in Ref.\cite{EL}. The
experimentally measurable forward-backward asymmetry $A_{FB}(b)$ and the 
mixed forward-backwrad and left-right asymmetry $A^{FB}_{LR}(b)$ for the
$b$ quark can be expressed as $A_{FB}(b)=\frac{3}{4}A_eA_b$ and 
$A^{FB}_{LR}(b)=\frac{3}{4}A_b$, respectively. One can thus determine
the asymmetry parameter $A_b$ model-independently both by using the SLC
measured value of $A^{FB}_{LR}(b)=\frac{3}{4}(0.867\pm 0.035)$ and by using 
the LEP measured $A_{FB}(b)=0.0990\pm 0.0021$ together with the averaged value 
of the LEP and SLC measured lepton asymmetry parameter $A_l=0.1489\pm 0.0018$
for $A_e$. The former gives $A_b=0.867\pm 0.035$ and the latter gives 
$A_b=0.887\pm 0.022$. The averaged value is then $A_b=0.881\pm 0.019$ 
which is {\it almost $3\sigma$ below the SM prediction} listed in Table
I \cite{EL}.

From the above analyses, we see that there are still considerable deviations
between the experiments and tha SM predictions provided one
concentrates to certain well measured data and analyzes the data 
model-independently. Of course, the above discrepancies still cannot
lead to a definite conclusion of needing new physics beyond the SM.
Definite conclusion can only be made together with more future experiments, 
e.g. experiments at the LC. It is interesting to notice that the discrepancies 
are related to the $b$ quark in {\it the third family}. If the discrepancies 
really reflect new physics beyond the SM, it will not be surprising since the 
top quark, as the $SU(2)$ partner of the $b$ quark in the third family, is
the heaviest particle yet dicovered whose mass is close to the EWSB scale
$v=246$ GeV so that the third family is more sensitive to new physics in the 
EWSB sector than the first two families do. 

\null\vspace{1cm}
\begin{center}
{\bf III. The Higgs Boson}
\end{center}

\vspace{0.2cm}

\null\noindent
{\bf 1. Theoretical Aspect of the SM Higgs Field}

Although the SM is successful at present energies, it contains several
theoretically unsatisfactory features, especially its Higgs sector. 
\begin{itemize}
\item
There are so many free parameters in the SM and {\it most of them are related
to the Higgs filed}.
\item
The self-energy of the elementary Higgs filed theory is quadratically divergent
which makes the theory {\it unnatural} \cite{Chanowitz,Susskind} and will
lead to the {\it hierarchy} problem when grand unification is concerned.
\item
The renormalized coupling constant $\lambda$ of the Higgs self-interaction
vanishes in the continuum limit (i.e. when the regularization momentum
cut-off goes to infinity). This is the so-called {\it triviality}
problem \cite{Chanowitz}.
\end{itemize}

To avoid {\it triviality}, people usually take the point of view  that the SM 
may not be a fundamental theory but is a low energy effective
theory of a more fundamental theory below a certain physical scale $\Lambda$.
Since the Higgs mass $m_H$ is proportional to $\lambda$, there is an upper 
bound on the Higgs mass for a given value of $\Lambda$
\cite{HR}. Such a triviality bound on the Higgs mass is shown as the
upper curve in Fig. 1 \cite{HR}. By definition, $m_H$ cannot exceed $\Lambda$.
This determines the maximal value of $m_H$ which is {\it of the order of 1
TeV} \cite{Chanowitz,HR}.

\null\vspace{0.4cm}

\centerline{\epsfig{figure=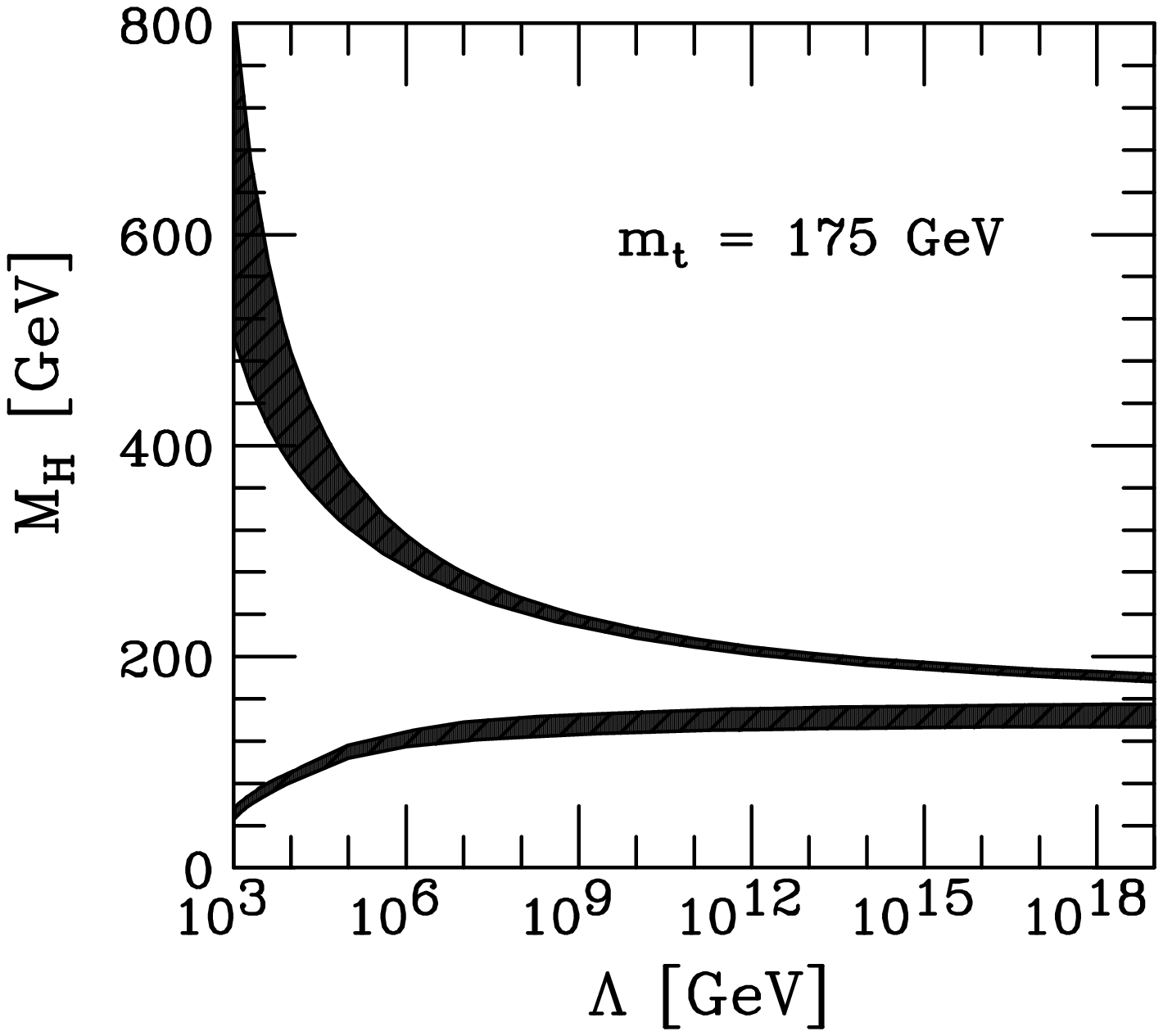,height=6cm,angle=90}}
\vspace{0.4cm}
\fcaption{The triviality bound (upper curve) and the vacuum
stability bound (lower curve) on $m_H$ in the SM.
The solid areas as well as the cross-hatched area indicate theoretical 
uncertainties. Quoted from Ref.\cite{HR}.} 
\null\vspace{0.2cm}

On the other hand, the Higgs boson 
self-interaction makes the physical vacuum a stable ground state with 
nonvanishing vacuum expectation value (VEV), while the fermion-loop 
contribution to the effective potential tends to {\it violate the vacuum 
stability}. The heavier the fermion is, the stronger the violation of vacuum 
stability will be. The $t$ quark is the heaviest fermion yet found which gives 
a strong violation of vacuum stability. Therefore we need a strong enough 
Higgs self-interaction (large enough $m_H$) to overcome the violation effect 
from the $t$ quark loop and maintain vacuum stability. This requirement gives 
a lower bound on the Higgs mass $m_H$. The vacuum stability bound on $m_H$ is
shown as the lower curve in Fig. 1 \cite{HR}.

We see from Fig. 1 that the scale of new physics $\Lambda$ can be of
the order of the Planck mass only if $m_H\sim 160$ GeV. Of special
interest is that if a very light Higgs boson with $m_H\sim 100$ GeV or
a heavy Higgs boson with $m_H\gtrsim 500$ GeV is found, the scale of new 
physics will be of the order of TeV. If the Higgs boson is not found below 1
TeV, we should find new physics beyond the SM in this region.

\null\vspace{0.2cm}
\null\noindent
{\bf 2. Searching for the Higgs Boson at High Energy Colliders}

 In this subsection, we briefly review the searches for the Higgs boson at high
energy colliders. For technical details, we refer to the talk by S. Yamashita 
at this workshop.

\null\noindent
{\bf 2.1 The CERN LEP2} \cite{LEP2}

At the LEP2 energy, the dominant production mechanism for the SM Higgs boson 
is the Higgs-strahlung process
\begin{eqnarray}
e^+e^-\to Z^*\to Z~H\,,
\label{Higgs-strahlung}
\end{eqnarray}
in which the Higgs boson is emitted from a virtual Z boson. The cross
section of this channel is nearly two orders of magnitude larger than
that of the W fusion of Higgs except at the edge of the phase space for 
Higgs-strahlung where both cross sections are very small. In the range of
$~50~{\rm GeV}<m_H<120~{\rm GeV}$, the main decay mode of the SM Higgs
boson is the $b\bar{b}$ mode [$B(H\to b\bar{b})\sim (80-90)\%$]. Other
branching ratios are smaller by an order of magnitude or more.
Experimental detections are in the following four modes\\
(a) $Z\to e^+e^-;\mu^+\mu^-,~~~~~~ H\to anything$\\
(b) $Z\to \tau^+\tau^-,~~~~~~~~~~~~~~H\to hadrons,$~and vice versa\\
(c) $Z\to \nu\bar{\nu},~~~~~~~~~~~~~~~~~H\to hadrons$\\
(d) $Z\to hadrons,~~~~~~~~~H\to b\bar{b}$.\\
Main detection backgrounds are fermion-pair productions, single
$W$ and $Z$ productions and $W,Z$ pair productions. For eliminating the
background $e^+e^-\to ZZ$ when $m_H\sim M_Z$, the $b$-tagging technique
is helpful since $B(Z\to b\bar{b})$ is only $(15.46\pm0.14)\%$ \cite{EL}. The
averaged signal to background ratio at the ALEPH, DELPHI, L3 and OPAL 
detectors is $(302\pm 7):(217\pm 11)$ \cite{LEP2}. At present
no evidence is found. The lower bound on $m_H$ at LEP2 is now
 \cite{DELPHI}.
\begin{eqnarray}
m_H>94.1~ GeV\,.
\end{eqnarray}

\null\noindent
{\bf 2.2 The Upgraded Fermilab Tevatron} 

In a recent paper \cite{HZ}, it is shown that the SM Higgs boson in the
mass region $135~{\rm GeV}\alt m_H\alt 180~{\rm GeV}$ is able to be detected
at the upgraded Tevtron with the center-of-mass energy of 2 TeV and an
integrated luminosity of 30 fb$^{-1}$ via the process
\begin{eqnarray}
p\bar{p}\to gg\to H\to W^*W^*\to l\nu jj\,,~~l\bar{\nu}\bar{l}\nu\,.
\end{eqnarray}
For the details of the signal and background analysis, see Ref.\cite{HZ}.

\null\noindent
{\bf 2.3 The CERN LHC}
\begin{itemize}
\item {\bf $m_H>140$ GeV}\\
\null\hspace{0.4cm}The clearest search is for $m_H>2M_Z$. In this case the following channel is 
available
\begin{eqnarray}
pp\to HX\to ZZX\to l^+l^-l^+l^-X~({\rm or}~l^+l^-\nu\bar{\nu}X)\,,
\end{eqnarray}
in which the four-lepton final state is very clear with rather small 
backgrounds. This channel is usually called {\it the golden channel}. 
Theoretical study shows that the resonance behavior can be
clearly seen when $m_H<800$~GeV \cite{Baur}.\\
\null\hspace{0.4cm}If $~140~{\rm GeV}<m_H<2M_Z$, the four-lepton final state can still be
detected with one of the $Z$'s virtual \cite{LHCW}.

\item {\bf $M_Z<m_H<140$ GeV}\\
\null\hspace{0.4cm}When the Higgs mass is in the intermediate range $M_Z<m_H<140~$GeV, the 
above detection is not possible since the branching ratio of the four-lepton 
channel drops very rapidly as $m_H<140~$GeV.
Detection of such an intermediate-mass SM Higgs boson is much more difficult. 
Fortunately, the $H\to \gamma\gamma$ branching ratio has its maximal
value in this $m_H$ range. Thus the best way is to detect the $\gamma\gamma$ 
final state for the Higgs boson. Recently, it is shown that the SM
Higgs boson in the mass range of 100 GeV$\--$150 GeV can be detected at
the LHC via
\begin{eqnarray}
pp\to H(\gamma\gamma)+jet
\end{eqnarray}
if a transverse-momentum cut of 2 GeV on the tracks is made for reducing the 
background \cite{KZ}. The statistical significance $S=N_S/\sqrt{N_B}$
($N_S$ and $N_B$ stand for the number of signal and number of
background, respectively) can be as large as $S>8$ for $100~{\rm
GeV}<m_H<150~{\rm GeV}$ \cite{KZ}.\\
\null\hspace{0.4cm}To find channels with larger rates and smaller
backgrounds, people suggested the following associate productions of $H$ 
\cite{A-P}.
\begin{eqnarray} 
&&pp\to WHX\to l\bar{\nu}\gamma\gamma X\,,\\
&&pp\to t\bar{t}HX\to l\bar{\nu}\gamma\gamma X\,.
\end{eqnarray}
In the $t\bar{t}H$ associate production channels, the leptons $l\bar{\nu}$ in 
the final state come from the $W$ decay in $t\to Wb$, thus it is an inclusive 
detection (without detecting $b$ and the decay products of $\bar{t}$). The 
signal and backgrounds of the $WH$ associate production channel have 
been calculated in Ref.\cite{A-P} which shows that the backgrounds are smaller 
than the signal even for a mild photon detector with a $3\%$ $\gamma\gamma$ 
resolution. The inclusive search for the $t\bar{t}H$ associate production 
suffers from a further large background from 
$pp\to W(\to l\bar{\nu})\gamma\gamma(n-jet),~(n=1,\cdots,4)$ \cite{ZK}, and 
the search is possible only when the $\gamma\gamma$ resolution of the photon 
detector is of the level of $1\%$ \cite{ZK}. The photon detectors of the CMS 
and ATLAS collaborations at the LHC are just of this level. Actually, if the 
jets are also detected, the background can be effectively reduced with certain 
choices of the jets, and such a detection is possible even for the mild photon 
detector with $3\%$ $\gamma\gamma$ resolution \cite{ZK}.
\end{itemize}

\newpage
\null\noindent
{\bf 2.4 The LC}

At the LC, the Higgs boson can be produced either by the
Higgs-strahlung process (\ref{Higgs-strahlung}) or by $WW$ and $ZZ$ fusions
\begin{eqnarray}
&&e^+e^-\to \nu\bar{\nu}(WW)\to \nu\bar{\nu}H\,,\\
&&e^+e^-\to e^+e^-(ZZ)\to e^+e^-H\,.
\end{eqnarray}
The cross sections for the Higgs-strahlung and $WW$ fusion processes are
$\sigma\sim 1/s$ and $\sigma\sim \frac{\ln\frac{s}{M_W}}{M^2_W}$, 
respectively.
So that the Higgs-strahlung process is important at $\sqrt{s}\alt 500$
GeV (e.g. the KEK JLC), while the $WW$ fusion process is important
at $\sqrt{s}>500$ GeV (e.g. the DESY TESLA and the SLAC NLC). Since there are 
less hadronic backgrounds at the LC, the Higgs boson can be tagged via the
$H\to b\bar{b}$ channel. Several thousands of events can be produced
for the envisaged luminosities \cite{LCPhys}.

By means of laser back-scattering, $\gamma\gamma$ and $e\gamma$ colliders can 
be constructed based on the LC. It has been shown recently that the 
$s$-channel Higgs production rate at the photon collider will be about an 
order of magnitude larger than the production rate in the Higgs-strahlung
process at the LC \cite{Telnov}.

It is shown that in the minimal SUSY extension of the SM, the cross
section of 
\begin{eqnarray}
\gamma\gamma\to hh
\end{eqnarray}
can be significantly enhanced if there is large $\tilde{t}$-mixing \cite{ZLG}, 
so that the lightest SUSY Higgs boson $h$ can be distinghuished from
that in the usual two-Higgs-doublet model through this process.

It has also been shown that at the $e\gamma$ collider, the process
\begin{eqnarray}
e\gamma\to e h
\end{eqnarray}
is enhanced by an order of magnitude by the chargino-loop contributions. So 
that this is a plausible process for detecting the lightest SUSY Higgs boson
$h$ \cite{L-R}.

The study of the $~f\bar{f}\gamma~$ decay mode of the Higgs bosons shows that,
for large $\tan\beta$ (say $\tan\beta\sim 30$), the decay rates of the
SUSY Higgs bosons $~h,~H,~A\to f\bar{f}\gamma~$ are significantly
smaller than that of the SM Higgs boson \cite{LZQ}, so that the SUSY
Higgs bosons can be distinghuished from the SM Higgs boson via this
decay mode.
 
\null\noindent
{\bf 2.5 The Muon Collider}

In recent years, there is an increasing attention to the construction of
muon colliders which is specially advantageous in detecting the SM
and the SUSY Higgs bosons in the $s$-channel \cite{MC}.
\begin{eqnarray}
\mu^+\mu^-\to H\to b(t)\bar{b}(\bar{t})\,.
\end{eqnarray}
If a light Higgs boson can be discovered at other colliders, e.g. the LHC or 
LC, a muon collider can be set up with energy around $m_H$ which will be 
specially useful for a better determination of $m_H$ and a measurement of
the total width $\Gamma^{tot}_H$ by scanning around $\sqrt{s}=m_H$ if the
energy spread is small enough. Otherwise, one can make a wider scan at
the muon colliders to search for the Higgs boson.
For $~m_H\leq 2M_W$, the branching ratio $B(H\to \mu^+\mu^-)$ is not
small due to the fact that the important decay channel $H\to WW,ZZ$ are
not open, so that this $s$-channel detection is very useful. In this case, 
the important decay channels can be $b\bar{b},~WW^*,~{\rm and}~ZZ^*$. The 
$WW^*$ channel can be measured via the final states $l\nu 2j$, and the 
$ZZ^*$ channel can be measured via $2l2j,~2\nu 2j,~4l~{\rm
and}~2l2\nu$. Then without knowing the precise value of $\Gamma(H\to \mu^+
\mu^-)$, one can measure the ratios of the $H\to b\bar{b}, ~H\to WW^*~
{\rm and}~H\to ZZ^*$ rates, and from which one can obtain the
$(Hb\bar{b})^2,~(HWW^*)^2~{\rm and}~(HZZ^*)^2$ coupling-squared ratios 
to the precision of a few percent \cite{MC}, which can test the Higgs boson 
interactions.
Higher energy muon colliders can also be used for the discovery of heavier SM
Higgs boson and the SUSY Higgs bosons $H$ and $A$ \cite{MC}. More on
the muon collider physics can be found in Ref.\cite{MC}

\newpage
\null\noindent
{\bf 3. Probing the Higgs Interactions}
 
If a light Higgs resonace is found from the above searches, it is {\it not 
the end of the story}. It is needed to test whether it is the SM Higgs or 
something else (for instance, a composite Higgs boson in certain
strongly interacting EWSB models). 
This can be done by examining its interactions. We know the
self-interactions of the SM Higgs boson contain the following trilinear
and quartic terms
\begin{eqnarray}
\frac{1}{8}\frac{m_H^2}{v^2}\bigg[4vH^3+H^4\bigg]\,,
\end{eqnarray}
where $v$ is the VEV of the Higgs field, $v=246$ GeV.
For detecting the trilinear interaction, it is possible to look at the
double Higgs-boson productions, i.e. $pp\to HHX$ at the LHC and
$e^+e^-\to HHZ,~~HH\bar{\nu}_e\nu_e$ at the LC. It has been shown that
the detection at the LHC is almost not possible due to the large background 
\cite{LCPhys,KZ}, while the detection at the LC is possible at the
C.M. energy $E=1.6$ TeV requiring a very large integrated luminosity,
$\int{\cal L}dt=1000$ fb$^{-1}$. Therefore the detection is not easy.
The signals of the quartic interactions are so small that it is hard to
detect. So that other methods of distinghushing the SM Higgs boson from
other Higgs resonances are needed. 

Since the top quark has the largest Yukawa coupling to the SM Higgs boson,
it is possible to detect the Higgs Yukawa coupling via the process
\begin{eqnarray}
e^+e^-\to t\bar{t}H\,.
\end{eqnarray}
This detection has been studied in Refs.\cite{Dawson,ttH}.

\null\vspace{1cm}
\begin{center}
{\bf IV. Supersymmetry}
\end{center}

The way of avoiding the unnaturalness problem without giving up the elementary 
Higgs fields is to introduce SUSY in which the
quardratic divergence of the Higgs self-energy causing unnaturalness is
canceled by the contribution of its SUSY partner. Furthermore, inclusion of 
the SUSY partners of known particles also weakened the problem of triviality. 
In most SUSY models motivated by avoiding the unnaturalness problem, the mass 
of the lightest Higgs boson $h$ is rather low, say below $130~$GeV \cite{HHW}.
Thus it is weakly interacting and is perturbatively calculable \footnote{It is 
still possible that there are other non-minimal heavy Higgs bosons which are 
strongly interacting} . 
SUSY theory is one of the promising theories of new physics which
receives most attention at present. 
However, it contains even more free parameters than the SM does. The number of 
free parameters may be reduced when going up to grand unification or even to 
superstring theory. Anyway, the most convincing test of SUSY models is to find 
SUSY partners experimentally. In this talk, we only briefly review some
processes of searching for SUSY particles. For details, we refer to the
talk given by K. Fujii at this workshop.

In the minimal SUSY extension of the SM (MSSM) \cite{MSSM}, both SUSY and 
anomaly-free require that the MSSM should contain two Higgs doublets 
$H_1$ and $H_2$. Every known particle and its SUSY partner (denoted by the 
corresponding symbol with a tilde) belong to a SUSY multiplet described
by a superfield (denoted by a symbol with a hat). To enforce lepton and baryon 
number conservation, a discrete symmetry called $R$-parity is usually imposed
\begin{eqnarray}
R=(-1)^{2s+3B+L}\,,
\end{eqnarray}
where $s$ is the spin quantum number. Then ordinary particles have
$R=1$ and SUSY particles have $R=-1$. The $R$-parity conserving
superpotential related to the Higgs sector can be written as \cite{MSSM}
\begin{eqnarray}
W=\epsilon_{ij}[(h_L)_{mn}\hat{H}_1^i\hat{L}_m^j\hat{E}_m
+(h_D)_{mn}\hat{H}_1^i\hat{Q}_m^j\hat{D}_n
-(h_U)_{mn}\hat{H}_2^i\hat{Q}_m^j\hat{U}_n-\mu\hat{H}_1^i\hat{H}_2^j]\,,
\label{SP}
\end{eqnarray}
where
$\hat{H}_1,~\hat{H}_2,~\hat{L}_m,~\hat{E}_n,~\hat{Q}_m,~\hat{D}_n,~{\rm
and}~\hat{U}_n$ are, respectively, superfields for $H_1,~H_2$, the left-habded 
lepton,  the right-handed lepton, the left-handed quark, the right-handed 
$D$- and $U$-quark, $h$'s are Yukawa couplings, and $\mu$ is a Higgs superfield
mass parameter. Since the real world does not respect SUSY
exactly, SUSY should be broken. In order to keep the solution of the 
unnaturalness problem, the SUSY breaking mechanism should not cause uncanceled
quadratic divergences so that it should be {\it soft}, and the 
scale of SUSY breaking should not be much higher than the TeV scale. With 
$R$-parity conservation, the general soft-SUSY-breaking terms are \cite{MSSM}
\begin{eqnarray}
V_{\rm soft}&=&m_1^2|H_1|^2+m_2^2|H_2|^2-m_{12}^2(\epsilon_{ij}H_1^iH_2^j+h.c.)
\nonumber\\
&&+(M_{\tilde{Q}}^2)_{mn}\tilde{Q}_m^{i*}\tilde{Q}_n^i
+(M_{\tilde{U}}^2)_{mn}\tilde{U}_m^*\tilde{U}_n
+(M_{\tilde{D}}^2)_{mn}\tilde{D}_m^*\tilde{D}_n
+(M_{\tilde{L}}^2)_{mn}\tilde{L}_m^{i*}\tilde{L}_n^i
+(M_{\tilde{E}}^2)_{mn}\tilde{E}_m^*\tilde{E}_n
\nonumber\\
&&+\epsilon_{ij}[(h_LA_L)_{mn}\tilde{H}_1^i\tilde{L}_m^j\tilde{E}_n
+(h_DA_D)_{mn}\tilde{H}_1^i\tilde{Q}_m^j\tilde{D}_n
-(h_UA_U)_{mn}\tilde{H}_2^i\tilde{Q}_m^j\tilde{U}_n+h.c.]
\nonumber\\
&&+\frac{1}{2}[M_3\tilde{g}\tilde{g}+M_2\tilde{W}^a\tilde{W}^a
+M_1\tilde{B}\tilde{B}+h.c.]\,,
\label{soft}
\end{eqnarray}
where the coefficients are complex matrices. The total number of
free parameters in the MSSM is 124 (105 new parameters in addition to
the 19 SM parameters) \cite{MSSM}. It is hard to make phenomenological 
predictions with so many free parameters. Actually, requirement on separate 
conservation of the lepton numbers $L_e,~L_\mu,~L_\tau$ and the experimental 
bounds on the flavor changing neutral current and the electric dipole moment of
the electron and neutron restrict the MSSM pareameters to a limited subspace
of the total parameter space. There are two general approaches which make the 
MSSM phenomenologically viable.
\begin{itemize}
\item {\bf The minimal supergravity-inspired MSSM (constrained MSSM)}.\\
\null~~In this approach the scalar squared masses and the $A$'s are flavor
diagonal and universal at the Planck scale $M_P$
\begin{eqnarray}
&&M^2_{\tilde{Q}}(M_P)=M^2_{\tilde{U}}(M_P)=M^2_{\tilde{D}}(M_P)
=M^2_{\tilde{L}}(M_P)=M^2_{\tilde{E}}(M_P)=m_0^2~{\bf 1}\,,\nonumber\\
&&m_1^2(M_P)=m_2^2(M_P)=m_0^2\,,\nonumber\\
&&A_U(M_P)=A_D(M_P)=A_L(M_P)=A_0~{\bf 1}\,,
\end{eqnarray}
and, furthermore, the gauge couplings and the gaugino mass parameters are 
assumed to unify at the grand unification scale $M_X$
\begin{eqnarray}
&&\sqrt{5/3}~g_1(M_X)=g_2(M_X)=g_3(M_X)=g_U\,,\nonumber\\
&&M_1(M_X)=M_2(M_X)=M_3(M_X)=m_{1/2}\,.
\end{eqnarray}
The total number of free parameters in this approach reduces to 23,
i.e. the 18 SM parameters (not including the Higgs mass) and 5 additional
new parameters $m_0,~m_{1/2},~A_0,~\tan\beta$ (the ratio of the VEV's
of the two Higgs bosons) and sgn($\mu$). This approach is also called the
{\it constrained MSSM} (CMSSM).
\item {\bf Models of gauge-mediated SUSY breaking}.\\
\null~~This kind of approach consists of a hidden sector in which SUSY is
broken, a "messenger" sector containing messenger fields with
$SU(3)\times SU(2)\times U(1)$ quantum numbers, and a sector containing
the fields of the MSSM. the coupling of the messengers to the hidden
sector generates a SUSY-breaking spectrum in the messenger sector, and
the SUSY breaking is transformed to the MSSM via the virtual exchange
of the messengers. In the simplest case of the models, there is one
effective mass scale $\Lambda$ which determines all the low-energy
scalar and gaugino masses though loop-effects (no $A$-parameters are 
generated). One must have $\Lambda\sim 100$ TeV in order to result the
SUSY partner masses of $O(1)~{\rm TeV}$ or less.The generation of $\mu$
and $m_{12}$ lies outside the ansatz of this SUSY breaking. There is a
messenger scale $M$ characterizing the average mass of the messenger particles
which can lie between $\Lambda$ and $10^{16}$ GeV. The total number of
free-parameters in this approach is 22, i.e. the 18 SM parameters (not
including the Higgs mass), and the four additional new parameters $\Lambda,~
\tan\beta,$ sgn($\mu$), and $M$. This simplest approach is called the
{\it minimal gauge-mediated model} (MGM).
\end{itemize}

Different approaches give rise to different SUSY particle spectra.
Particles with the same $SU(3)\times U(1)_{em}$ quantum numbers may mix
to form mass eigen-states. For example, the charginos
$\tilde{\chi}_j^{\pm},~~(j=1,2)$ are linear combinations of the charged
winos and higgsinos, and the neutralinos
$\tilde{\chi}_k^0~~(k=1,\cdots,4)$ are linear combinations of the
neutral wino, bino and neutral higgsinos. With $R$-parity conservation,
SUSY particles ($R=-1$) can only be produced in pairs, and heavy SUSY
particles ($R=-1$) cannot decay into ordinary particles ($R=1$) but
will decay eventually to the {\it lightest SUSY particle} (LSP) which behaves 
as a missing energy $E_T\!\!\!\!\!\!\!/\,\,\,\,$ in the experiments.
These characterize the phenomenology of SUSY searches. In the
following, we briefly review some of the SUSY searches at high energy 
colliders.
\begin{itemize}
\item {\bf LEP2}\\
\null~~Searching for SUSY particles has been carried out at LEP for many years,
and no evidence is found. We quote here some recent gives the following
lower bounds of the SUSY particles given in the  $\sqrt{s}=183$ GeV run. 

~~The DELPHI data on $e^+e^-\to \tilde{\chi}^0_1\tilde{\chi}^0_1\to 
\tilde{G}\gamma\tilde{G}\gamma$ leads to the lower bound of 
$m_{\tilde{\chi}^0_1}$ in the MGM assuming $\tilde{\chi}^0_1\to
\tilde{G}\gamma$ to have $100\%$ probability and negligible lifetime
\cite{DELPHI98}
\begin{eqnarray}
m_{\tilde{\chi}^0_1}>83~{\rm GeV}
\end{eqnarray}
with $m_{\tilde{e}_R}=1.1m_{\tilde{\chi}^0_1}$ and
$\tilde{\chi}_1^0\approx \tilde{B}$.

~~The OPAL search for sleptons leads to the following bounds. For $\mu<-100$
GeV and $\tan\beta=1.5$, the $95\%$ C.L. bounds are \cite{OPAL}
\begin{eqnarray}
&&m_{\tilde{e}_R}<77~{\rm GeV~~~~for}~m_{\tilde{e}^-}-m_{\tilde{\chi}_1^0}
>5~{\rm GeV}\,,\nonumber\\
&&m_{\tilde{\mu}_R}<65~{\rm GeV~~~~for}~m_{\tilde{\mu}^-}
-m_{\tilde{\chi}_1^0}>2~{\rm GeV}\,,\nonumber\\
&&m_{\tilde{\tau}_R}>60~{\rm
GeV~~~~for}~m_{\tilde{\tau}^-}-m_{\tilde{\chi}_1^0}>9~{\rm GeV}\,.
\end{eqnarray}

It has been shown that the LEP precision measurements at the $Z$-pole
and low-energy electroweak experiments give constraints on the MSSM
parameters, especially the $b\to s\gamma$ data gives severe constraint
on the parameter space with $\mu>0$ and large $\tan\beta$ \cite{EP}.

\item {\bf LHC}\\
\null~~At the LHC, SUSY particles pairs $\tilde{g}\tilde{g},~\tilde{g}\tilde{q},
~\tilde{q}\tilde{q},~\tilde{l}\tilde{l},~\tilde{\chi}_1^0\tilde{\chi}_1^\pm$
can be searched for via the following channels \cite{Kunszt}
\begin{eqnarray}
&& l^\pm+jets+E_T\!\!\!\!\!\!\!/\,\,\,\,:~~~~
\tilde{g}\tilde{g},~\tilde{q}\tilde{q}\\ \nonumber
&& l^\pm l^\pm +jets+E_T\!\!\!\!\!\!\!/\,\,\,\,:~~~~
\tilde{g}\tilde{g},~\tilde{g}\tilde{q},
~\tilde{q}\tilde{q}\\ \nonumber
&& l^\pm l^\pm l^\pm +E_T\!\!\!\!\!\!\!/\,\,\,\,:~~~~
\tilde{\chi}^0_1\tilde{\chi}^\pm_1\\ \nonumber
&& l^\pm l^\pm +E_T\!\!\!\!\!\!\!/\,\,\,\,:~~~~\tilde{l}\tilde{l}\,.
\label{LHCSUSY}
\end{eqnarray}
The LHC is advantageous for detecting colored SUSY particles. For
example, in the first channel in (\ref{LHCSUSY}), $\tilde{g}$ and $\tilde{q}$ 
can be detected up the mass values of 3.6 TeV \cite{Kunszt}.

\item {\bf LC}\\
\null~~The LC is advantageous due to the small backgrounds. It provides
high precision detections of all SUSY particles up to the mass values
of 1 TeV \cite{LCPhys}. 

~~Neutralinos and charginos are easy to detect and study with high accuracy at 
the LC via \cite{LCPhys}
\begin{eqnarray}
&& e^+e^-\to \tilde{\chi}^+_i\tilde{\chi}^-_j~~~~,i,j=1,2\\ \nonumber
&& e^+e^-\to \tilde{\chi}^0_i\tilde{\chi}^0_j~~~~,i,j=1,2,3,4
\end{eqnarray}
with $\tilde{\chi}_1^0\tilde{\chi}_1^0\to
\tilde{G}\tilde{G}\gamma\gamma\to \gamma\gamma E_T\!\!\!\!\!\!\!/\,\,\,\,$.
The neutralino and chargino masses can be measured to the accuracy of
$\delta m_{\tilde{\chi}^\pm_1}\approx 100~$MeV,~
$\delta m_{\tilde{\chi}^0_1}\approx 600~$MeV.

~~The sleptons can be detected and studied via \cite{LCPhys}
\begin{eqnarray}
&& e^+e^-\to \tilde{l}^+_L\tilde{l}^-_L,~\tilde{l}^+_R\tilde{l}^-_R,~{\rm etc.\,,}
\\ \nonumber
&& e^+e^-\to \tilde{t}_1\tilde{t}_1,~\tilde{t}_2\tilde{t}_2,~{\rm etc.}
\end{eqnarray}
The smuon mass can be measured to the accuracy of 
$\delta m_{\tilde{\mu}}\approx 1.8~$GeV.

~~It has been shown that the cross sections of $~e^-e^-\to \tilde{e}^-_{L,R}
\tilde{e}^-_{L,R}~$ are much larger than those of 
$~e^+e^-\to \tilde{e}^+_{L,R}\tilde{e}^-_{L,R}~$ \cite{e-e-}, so that the
$e^-e^-$ linear collider is specially advantageous for the search for 
selectron.
\end{itemize}

We see that, with the LHC and the LC, it is possible to search for all the 
SUSY particles with the mass values up to 1 TeV which covers the range 
required by solving the unnaturalness problem by SUSY. If SUSY particles are
found below 1 TeV, SUSY will be confirmed and one should study the
properties of the SUSY particles seriously to see if the model is
just the simple MSSM or certain complicated SUSY model beyond the MSSM. If
SUSY particles are not found in this energy region, SUSY will be
{\it irrelevant to the solution of the unnaturalness problem}, and we should
consider other possible solutions, e.g. strongly interacting
EWSB mechanism (without introducing elementary
Higgs fields) which will be reviewed in the next section.

Another interesting feature is that the careful theoretical study on the MSSM 
Higgs mass up to two-loop calculations show that the mass of the lightest CP 
even Higgs boson $h$ in the MSSM cannot exceed a bound $m_h|_{max}\approx 130$ 
GeV \cite{HHW}. Thus the Higgs boson $h$ can be searched for at all the 
designed LC's (NLC, TESLA and JLC). If a light scalar resonance is found below
130 GeV, it will be a good support to the MSSM, and further study of its 
properties will be needed to see if it is really the MSSM Higgs boson.
If no light scalar resonance is found below 130 GeV, MSSM
will be in a bad shape and SUSY models beyond the MSSM should be
seriously considered.

\null
\vspace{1cm} 
\begin{center}
{\bf V. Strongly Interacting Electroweak Symmetry Breaking Mechanism}
\end {center}

\vspace{0.3cm}
\null\noindent
{\bf 1. Strongly Interacting Electroweak Symmetry Breaking Models}

Introducing elementary Higgs field is the simplest but not unique way
of breaking the electroweak gauge symmetry spontaneously. The way of completely 
avoiding the problems of triviality and unnaturalness is to abandon elementary
scalar fields and introducing new strong interactions causing certain fermion 
condensates to break the electroweak gauge symmetry. This idea is similar to 
those in the theory of superconductivity and chiral symmetry breaking
in QCD. The simplest model realizing this idea is the original QCD-like
technicolor model \cite{Weinberg,Susskind}. However, such a simple model 
predicts a too large oblique correction parameter $S$ \cite{STU} and is already
ruled out by the LEP data. A series of improved models have been
proposed to overcome the shortcomings of the simplest model 
\cite{WTC,AT,MTC,TC2,NCETC,LR,topseesaw}. In the following, we briefly review 
two of the recently proposed models.
\begin{itemize}
\item Topcolor-Assisted Technicolor Models\\
\null~~This model combines the technicolor idea and the top-condensate idea,
in which a large enough $m_t$ and $m_b$ mass difference can be obtained
without causing large oblique correction parameters $S,~T$ and $U$ 
\cite{TC2}. It is assumed in this model that at the energy scale
$\Lambda\gtrsim 1$ TeV, there is a {\it topcolor} theory with the gauge 
group $SU(3)_1\times U(1)_{Y1}\times SU(3)_2\times U(1)_{Y2}\times SU(2)_L$
in which $SU(3)_1\times U(1)_{Y1}$ preferentially couples to the third-family
fermions and $SU(3)_2\times U(1)_{Y2}$ preferentially couples to the first- 
and second-family fermions, so that the third-family fermions are
diffrerent from those in the first two families. It is assumed that there is
also a technicolor sector mainly in charge of the EWSB and will break the 
topcolor gauge group into $SU(3)_{QCD}$ and $U(1)_Y$ at the scale $\Lambda$:
\begin{eqnarray}
SU(3)_1\times U(1)_{Y1}\times SU(3)_2\times U(1)_{Y2}\times SU(2)_L\to
SU(3)_{QCD}\times U(1)_Y\times SU(2)_L\,.
\end{eqnarray}
The $SU(3)_1\times U(1)_{Y1}$ couplings are assumed to be much stronger
than those of $SU(3)_2\times U(1)_{Y2}$. The strong $SU(3)_1\times U(1)_{Y1}$
interactions will form top quark condensate $\langle t\bar{t}\rangle$ but not 
bottom quark condensate from the simultaneous effects of the $SU(3)_1$ and
$U(1)_{Y1}$ interactions. The technicolor dynamics gives rise to the
masses of the $u,~d,~s,~c,~$ and $b$ quarks and a small portion of the
top quark mass, while the main part of the top quark mass comes from
the topcolor dynamics causing the top quark condensate just like the
constituent quarks acquiring their large dynamical masses from the dynamics
causing the quark condensates in QCD. In this prescription, the
technicolor dynamics does not cause a large oblique correction
parameter $T$. Improvement of this kind of model is still in progress 
\cite{Lane}. It has been shown that this kind of model gives rise to a positive
enhancement to $R_b$ with the right order of magnitude, which depends on the 
parameters in the model \cite{TCRB}.

~~This kind of model contains the usual pseudo-Goldstone bosons (PGB's)
in the techicolor sector with masses of few hundred GeV and an isospin
triplet top-pions with masses around 200 GeV. These light particles
characterizing the phenomenology of the model.

\item Top Quark Seesaw Theory\\
\null~~Very recently, a new promising theory of strongly interacting 
EWSB related to the top quark condenstate
called {\it top quark seesaw theory} was proposed \cite{topseesaw}. The
gauge group in this theory is $SU(3)_1\times SU(3)_2\times
SU(2)_W\times U(1)_Y$ which breaks into $SU(3)_{QCD}\times U(1)_Y$ at a
scale $\Lambda$. Instead of introducing techniquarks, certain 
$SU(2)_W$-singlet quarks, $\chi,\cdots$, with topcolor interactions and
specially assigned $U(1)_Y$ quantum numbers are introduced in 
this theory. Topcolor causes the following $t$ ($b$) and $\chi$ bound state
scalar field
\begin{eqnarray}
\varphi=\pmatrix{\overline{\chi_R}~t_L\cr \overline{\chi_R}~b_L}
\end{eqnarray}
which behaves like a Higgs doublet whose VEV breaks the electroweak symmetry.
Furthermore, the VEV of $\varphi$ will cause a dynamical mass 
$m_{t\chi}\sim 600$ GeV, and the dynamics in this theory will cause the
following mass terms in the $\chi-t$ sector
\begin{eqnarray}
-\mu_{\chi\chi}\overline{\chi_L}\chi_R-\mu_{\chi t}\overline{\chi_L}t_R+h.c.
\end{eqnarray}
Then the mass matrix of the heavy charge $2/3$ quarks takes the form
\begin{eqnarray}
(\overline{t_L}~~\overline{\chi_L})\,
{\left(\begin{array}{cc}
0 & m_{t\chi} \\
\mu_{\chi t} & \mu_{\chi\chi}\\
\end{array} \right)}
\left( \begin{array}{c}
t_R\\
\chi_R\\
\end{array} \right)
\label{massmatrix}
\end{eqnarray}
which gives the seesaw mechanism for the top quark mass. Diagonalizing
the mass matrix in (\ref{massmatrix}) for $\mu_{\chi\chi}\gg m_{t \chi}$
leads to the physical top quark mass
\begin{eqnarray}
m_t\approx m_{t \chi}\frac{\mu_{\chi t}}{\mu_{\chi\chi}}\,.
\label{mt}
\end{eqnarray}
Eq.(\ref{mt}) can yield the desired top quark mass for appropriate dynamical
value of $\mu_{\chi t}/\mu_{\chi\chi}$.

~~The composite scalar spectrum in this theory is:
\begin{itemize}
\item $h^0$: neutral Higgs boson of mass $m_{t\chi}$ times a factor of order 
one (or smaller);
\item $H^0,~H^\pm,~A^0$: heavy states of a two Higgs-doublet sector (roughly 
degenerate);
\item $H^0_{\chi t},~A^0_{\chi t}$: a CP-even and a CP-odd state (can
be light);
\item $A^0_{\chi\chi}$: a CP-odd state (can be light);
\item $\phi_{bb}$: neutral complex scalar (with arbitrary mass);
\item $H^\pm_{tb}$: charged scalar which can be as light as 250 GeV;
\item $H^0_{\chi\chi},~H^\pm_{\chi b}$: CP-even neutral state and a
charged scalar with larges masses.
\end{itemize}

~~This theory has several advantages. (a) In this model, one of the particles 
reponsible for the EWSB is just the known top quark, and the $SU(2)_W$-doublet 
nature of the Higgs filed just comes from the $SU(2)_W$-doublet nature of the 
third family quarks. (b) The new quark $\chi$ introduced in this theory is 
$SU(2)_W$-singlet so that there is no large custodial symmetry violation, and 
the experimental constraint on the oblique correction parameter $T$ can be
easily satisfied. (c) The problem of predicting a too large oblique
correction parameter $S$ in technicolor theories due to introducing
many technifermion-doublets deos not exsist in the present theory
since there is only one top quark condensate. (d) Unlike
the original top quark condensate model which leads to a too large top
quark mass, the present theory can give rise to the desired top quark mass 
via the seesaw mechanism.

~~Various possible ways of building models in this theory are given
in Ref.\cite{topseesaw}.
\end{itemize}

\vspace{0.3cm}
\null\noindent
{\bf 2. Model-Dependent Tests}

Due to the nonperturbative nature of the strong interaction dynamics,
it is hard to make precision predictions from the strongly interacting
electroweak symmetry breaking models. However, most of the models
contain certain PGB's with masses in the region of few hundred GeV. The 
properties of the PGB's are diffrerent from model to model, therefore
the PGB's are characteristics of the models, and their effects can be
experimentally tested. Productions of PGB's, especially the PGB's in the 
technicolor sector, at the existing high energy colliders have been 
extensively studied in the literatures \cite{PGBprod,HY}. 

Since the top quark couples to the EWSB sector strongly due to its large mass, 
a feasible way of testing the strongly interacting EWSB models is to test the
PGB effects in top quark productions at high energy colliders. It is
shown that at the Tevatron \cite{ELtt}, the LHC \cite{ttLHC} and the
photon collider \cite{phphtt}, the $s$-channel PGB effects in top quark 
pair productions are experimentally detectable, and with which models with and 
without topcolor can be experimentally distinguished and can be
distinguished from the MSSM \cite{phphtt}.

\vspace{0.3cm}
\null\noindent
{\bf 3. Model-Independent Probe of the EWSB Mechanism}

We have seen that there are various kinds of EWSB models proposed. Some
of them contain light Higgs boson(s) (elementary or composite) and some
of them do not. We still do not know whether the actual EWSB mechanism
in the nature looks like one of the proposed models or not. Therefore,
only testing the proposed models seems to be not enough, and certain 
model-independent probe of the EWSB mechanism is needed. Since the
scale of new physics is likely to be several TeV, electroweak physics at energy
$E\lesssim 1$ TeV can be effectively described by the {\it electroweak
effective Lagrangian} in which composite fields are approximately
described by effective local fields. The electroweak effective Lagrangian is a 
{\it general description} (including all kinds of models) which contains 
certain yet unknown coefficients whose values are, in principle, 
determined by the underlying dynamics. Different EWSB models give rise to 
different sets of coefficients. The model-independent probe (the first step) 
is to investigate through what processes and to what accuracy we can measure 
these coefficients in the experiments \footnote{This is in a similar spirit as 
the study of the chiral Lagranian in QCD by Gasser and Leutwyler
\cite{GL}.} .
Once this is done, the second step is to examine what kind of model can give
rise to a set of coefficients fitting the experimental values. The
sencond step concerns the difficult strong interaction dynamics,
and we shall not discuss it in this short review article. We shall
mainly consider the first step.

From the experimental point of view, the most challenging case of probing
the EWSB mechanism is that there is no light scalar resonance found
below 1 TeV. We shall take this case as the example in this review.
Effective Lagrangian including a light Higgs boson has also been studied in the
literature \cite{Buchmuller}. In the case we are considering, the
effective Lagrangian is the so called electroweak chiral Lagrangian (EWCL)
which is an effective Lagrangian for the would-be Goldstone bosons $\pi^a$
in the nonlinear realization $U=e^{i\tau^a\pi^a/f_\pi}$ with
electroweak interactions. The bosonic sector of which, up 
to the $p^4$-order, reads \cite{AW,et4}
\begin{eqnarray}
&&{\cal L}_{\rm eff}={\cal L}_{\rm G}+{\cal L}_{\rm S}\,,
\label{CL1}
\end{eqnarray}
with
\begin{eqnarray}
&&{\cal L}_{\rm G}=-\frac{1}{2}{\rm Tr}({\bf W}_{\mu\nu}{\bf W}^{\mu\nu})
-\frac{1}{4}B_{\mu\nu}B^{\mu\nu}  ~~,\nonumber\\
&&{\cal L}_{\rm S}= {\cal L}^{(2)}+{\cal L}^{(2)\prime}+
             \displaystyle\sum_{n=1}^{14} {\cal L}_n ~~,
\label{CL2}		   
\end{eqnarray}
where ${\bf W}_{\mu}\equiv W^a_{\mu}\displaystyle\frac{\tau^a}{2}~,~~~
{\bf B}_{\mu}\equiv B_{\mu}\displaystyle\frac{\tau^3}{2}~,~~~
{\bf W}_{\mu\nu} =\partial_{\mu}{\bf W}_{\nu}-\partial_{\nu}{\bf W}_{\mu}
                      +ig [{\bf W}_\mu , {\bf W}_\nu ] ~,~~~
B_{\mu\nu}= \partial_\mu B_\nu - \partial_\nu B_\mu~,$ and          
\vspace{0.2cm}
\begin{eqnarray}
&&{\cal L}^{(2)}=\frac{f_\pi^2}{4}{\rm Tr}[(D_{\mu}U)^\dagger(D^{\mu}U)]~~,
~~~~~~~~~~~~~~
{\cal L}^{(2)\prime} =\ell_0 (\frac{f_\pi}{\Lambda})^2~\frac{f_\pi^2}{4}
               [ {\rm Tr}({\cal T}{\cal V}_{\mu})]^2 ~~,\nonumber\\
&&{\cal L}_1 = \ell_1 (\frac{f_\pi}{\Lambda})^2~ \frac{gg^\prime}{2}
B_{\mu\nu} {\rm Tr}({\cal T}{\bf W^{\mu\nu}}) ~~,~~~~~~~~
{\cal L}_2 = \ell_2 (\frac{f_\pi}{\Lambda})^2 ~\frac{ig^{\prime}}{2}
B_{\mu\nu} {\rm Tr}({\cal T}[{\cal V}^\mu,{\cal V}^\nu ]) ~~,\nonumber\\
&&{\cal L}_3 = \ell_3 (\frac{f_\pi}{\Lambda})^2 ~ig
{\rm Tr}({\bf W}_{\mu\nu}[{\cal V}^\mu,{\cal V}^{\nu} ]) ~~,~~~~~~~~
{\cal L}_4 = \ell_4 (\frac{f_\pi}{\Lambda})^2 
                     [{\rm Tr}({\cal V}_{\mu}{\cal V}_\nu )]^2 ~~,\nonumber\\
&&{\cal L}_5 = \ell_5 (\frac{f_\pi}{\Lambda})^2 
                     [{\rm Tr}({\cal V}_{\mu}{\cal V}^\mu )]^2 ~~,
				 ~~~~~~~~~~~~~~~~~~
{\cal L}_6 = \ell_6 (\frac{f_\pi}{\Lambda})^2 
[{\rm Tr}({\cal V}_{\mu}{\cal V}_\nu )]
{\rm Tr}({\cal T}{\cal V}^\mu){\rm Tr}({\cal T}{\cal V}^\nu) ~~,\nonumber\\
&&{\cal L}_7 = \ell_7 (\frac{f_\pi}{\Lambda})^2 
[{\rm Tr}({\cal V}_\mu{\cal V}^\mu )]
{\rm Tr}({\cal T}{\cal V}_\nu){\rm Tr}({\cal T}{\cal V}^\nu) ~~,~~~~
{\cal L}_8 = \ell_8 (\frac{f_\pi}{\Lambda})^2~\frac{g^2}{4} 
[{\rm Tr}({\cal T}{\bf W}_{\mu\nu} )]^2  ~~,\nonumber\\
&&{\cal L}_9 = \ell_9 (\frac{f_\pi}{\Lambda})^2 ~\frac{ig}{2}
{\rm Tr}({\cal T}{\bf W}_{\mu\nu}){\rm Tr}
        ({\cal T}[{\cal V}^\mu,{\cal V}^\nu ]) ~~,~~~~
{\cal L}_{10} = \ell_{10} (\frac{f_\pi}{\Lambda})^2\frac{1}{2}
[{\rm Tr}({\cal T}{\cal V}^\mu){\rm Tr}({\cal T}{\cal V}^{\nu})]^2 ~~,\nonumber\\
&&{\cal L}_{11} = \ell_{11} (\frac{f_\pi}{\Lambda})^2 
~g\epsilon^{\mu\nu\rho\lambda}
{\rm Tr}({\cal T}{\cal V}_{\mu}){\rm Tr}
({\cal V}_\nu {\bf W}_{\rho\lambda}) ~~, ~~~~
{\cal L}_{12} = \ell_{12}(\frac{f_\pi}{\Lambda})^2 ~2g
                    {\rm Tr}({\cal T}{\cal V}_{\mu}){\rm Tr}
                  ({\cal V}_\nu {\bf W}^{\mu\nu}) ~~,\nonumber\\
&&{\cal L}_{13} = \ell_{13}(\frac{f_\pi}{\Lambda})^2~ 
      \frac{gg^\prime}{4}\epsilon^{\mu\nu\rho\lambda}
      B_{\mu\nu} {\rm Tr}({\cal T}{\bf W}_{\rho\lambda}) ~~,\nonumber\\
&&{\cal L}_{14} = \ell_{14} (\frac{f_\pi}{\Lambda})^2~\frac{g^2}{8} 
\epsilon^{\mu\nu\rho\lambda}{\rm Tr}({\cal T}{\bf W}_{\mu\nu})
{\rm Tr}({\cal T}{\bf W}_{\rho\lambda})~~,
\label{CL3}
\end{eqnarray}
\vspace{0.2cm}
in which $D_{\mu}U =\partial_{\mu}U + ig{\bf W}_{\mu}U 
-ig^{\prime}U{\bf B}_{\mu}~,~~~{\cal V}_{\mu}\equiv (D_{\mu}U)U^\dagger~$, 
~~and $~~{\cal T}\equiv U\tau_3 U^{\dagger}$. 

The coefficients $\ell$'s reflect the strengths of the $\pi^a$
interactions, i.e. the EWSB mechanism. $\ell_1,~\ell_0$ and $\ell_8$ are
related to the oblique correction parameters $S,~T$ and $U$,
respectively; $\ell_2,~\ell_3,~\ell_9$ are related to the 
triple-gauge-couplings;
${\cal L}_{12},~{\cal L}_{13}$ and ${\cal L}_{14}$ are CP-violating.
The task now is to find out experimental processes to measure the yet 
undetermined $\ell$'s \footnote{Note that with $\ell_1,~\ell_0,~\ell_8$ taking
the values from the experimentally values of $S,~T$ and $U$, 
$\ell_2,~\ell_3,~\ell_9$ consistent with the experimental bounds on the
triple-gauge-couplings which are still rather weak, and a
suitable $Zb\bar{b}$-coupling, the theory described by this chiral
Lagrangian (without a light Higgs boson) can fit the LEP precision
electroweak data.} . 

Note that the would-be-Goldstone bosons $\pi^a$ are not physical
particles, so that they are not experimentally observable. However, due to the 
Higgs mechanism, the degrees of freedom of $\pi^a$ are related to the 
longitudinal components of the weak bosons $V^a_L$ ($W^\pm_L,~Z^0_L$) which 
are experimentally observable. Thus the $\ell$'s are able to be
measured via $V^a_L$-processes. So that we need to know the quatitative 
relation between the $V^a_L$-amplitude (related to the experimental data) and 
the GB-amplitude (reflecting the EWSB mechanism), which is the so-called 
{\it equivalence theorem} (ET). ET has been studied by many papers 
\cite{et-tree}\cite{et1}\cite{et2}, and the final precise formulation
of the ET and its rigorous proof are given in 
Refs.\cite{et2}\cite{et3}\cite{et-new}. The precise formulation of the
ET is
\begin{eqnarray}                                                                             
T[V^{a_1}_L,V^{a_2}_L,\cdots]                                   
= C\cdot T[-i\pi^{a_1},\i\pi^{a_2},\cdots]+B ~~,
\label{ET}
\end{eqnarray}
with
\begin{eqnarray}
&&E_j \sim k_j  \gg  M_W , ~~~~~(~ j=1,2,\cdots ,n ~)~~,\nonumber\\
&&C\cdot T[-i\pi^{a_1},-i\pi^{a_2},\cdots]\gg B ~~,
\end{eqnarray}
where $~T[V^{a_1}_L,V^{a_2}_L,\cdots]~$ and 
$~T[-i\pi^{a_1},-i\pi^{a_2},\cdots]~$
are, respectively, the $V^a_L$-amplitude and the $\pi^a$-amplitude$, E_j$ is 
the energy of the $j$-th external line, $C$ is a gauge and
renormalization scheme dependent constant factor, and $B$ is a 
process-dependent function of the energy $E$. This precise formulation
has been proved both in the SM and in the EWCL formalism \cite{et2}. By taking 
special convenient renaormalization scheme, the constant $C$ can be
simplified to $C=1$ \cite{et2,et3,et-new}. In the EWCL theory, the
$B$-term may not be small even when the center-of-mass ernergy $E\gg M_W$, and 
{\it it is not sensitive to the EWSB mechanism}. Therefore the $B$-term serves 
as an {\it intrinsic background} when probing $~T[-i\pi^{a_1},-i\pi^{a_2},
\cdots]~$ via $~T[V^{a_1}_L,V^{a_2}_L,\cdots]$ in (\ref{ET}). Only when 
$~|B|\ll |C\cdot T[-i\pi^{a_1},-i\pi^{a_2},\cdots]|~$ the probe can be
{\it sensitive}. In Ref.\cite{global}, a new power counting rule for 
semi-quantitatively estimating the amplitudes in the EWCL theory was proposed, 
and with which a systematic analysis on the sensitivities of probing the EWSB
mechanism via the $V^a_L$ processes were given. The results are
summarized in Table II.

\newpage
\tabcolsep 1pt

\tcaption{ 
~{\small Probing the EWSB Sector at High Energy Colliders:
A Global Classification for the NLO Bosonic Operators.} 
Quoted from Ref.\cite{global}. \\[0.3cm]
 (~Notations: ~$\surd =~$Leading contributions, 
$~\triangle =~$Sub-leading contributions,~ 
and ~$\bot =~$Low-energy contributions.~
~Notes:~ $^{\dagger}$Here, $~{\cal L}_{13}$ or $~{\cal L}_{14}~$ 
does not contribute at $~O(1/\Lambda^2)~$.   ~~ $^\ddagger$At LHC($14$),  
$W^+W^+\ra W^+W^+$ should also be included.~)  }

\renewcommand{\baselinestretch}{1.5} 
\renewcommand{\arraystretch}{1.2}
\scriptsize
\vspace{0.3cm}

\begin{tabular}{||c||c|c|c|c|c|c|c|c|c|c||c||c||} 
\hline\hline
& & & & & & & & & & & & \\
 Operators 
& $ {\cal L}^{(2)\prime} $ 
& $ {\cal L}_{1,13} $ 
& $ {\cal L}_2 $
& $ {\cal L}_3 $
& $ {\cal L}_{4,5} $
& $ {\cal L}_{6,7} $ 
& $ {\cal L}_{8,14} $ 
& $ {\cal L}_{9} $
& $ {\cal L}_{10} $
& $ {\cal L}_{11,12} $
& $T_1~\parallel  ~B$ 
& Processes \\
& & & & & & & & & & & & \\
\hline\hline
 LEP-I (S,T,U) 
& $\bot$ 
& $\bot~^\dagger$
&  
& 
& 
& 
& $\bot~^\dagger$
& 
&
&
& $g^4\frac{f^2_\pi}{\Lambda^2}$ 
& $e^-e^+\ra Z \ra f\bar{f}$\\ 
\hline
  LEP-II
& $\bot$ 
& $\bot$  
& $\bot$  
& $\bot$  
&  
& 
& $\bot$  
& $\bot$ 
&
& $\bot$  
& $g^4\frac{f^2_\pi}{\Lambda^2}$
& $e^-e^+ \ra W^-W^+$\\
\hline
  LC($0.5$)/LHC($14$)
& 
& 
& $\surd$
& $\surd$
& 
& 
& 
& $\surd$
&
& 
& $g^2\frac{E^2}{\Lambda^2} \parallel g^2\frac{M_W^2}{E^2}$
& $f \bar f\ra W^-W^+ /(LL)$\\  
& 
& $\triangle$
& $\triangle$
& $\triangle$
& 
& 
& $\triangle$
& $\triangle$
&
& $\triangle$
& $g^3\frac{Ef_\pi}{\Lambda^2} \parallel g^2\frac{M_W}{E} $ 
& $f \bar f\ra W^-W^+/(LT) $\\  
\hline
& 
& 
& 
& $\surd$
& $\surd$
& $\surd$
& 
& $\surd$
&
& $\surd$
& $g^2\frac{1}{f_\pi}\frac{E^2}{\Lambda^2}
  \| g^3\frac{M_W}{E^2} $
& $f \bar f\ra W^-W^+Z /(LLL) $\\
& 
& $\triangle$ 
& $\triangle$
& $\triangle$
& $\triangle$
& $\triangle$
& $\triangle$
& $\triangle$
&
& $\triangle$
& $g^3\frac{E}{\Lambda^2}\parallel g^3\frac{M_W^2}{E^3}$ 
& $f \bar f\ra W^- W^+ Z /(LLT)  $\\
& 
& 
&  
& $\surd$
& $\surd$
& $\surd$
& 
& 
& $\surd$
& 
& $g^2\frac{1}{f_\pi}\frac{E^2}{\Lambda^2}\parallel 
  g^3\frac{M_W}{\Lambda^2}$
& $f \bar f \ra ZZZ /(LLL) $\\
& 
& 
& 
& 
& $\triangle$
& $\triangle$
& 
& 
& $\triangle$
& 
& $g^3\frac{E}{\Lambda^2}\parallel
   g^3\frac{f_\pi}{\Lambda^2}\frac{M_W}{E}$ 
& $f \bar f \ra ZZZ  /(LLT)  $\\
 ~LC($1.5$)/LHC($14$)~ 
& 
& 
& 
& 
& $\surd$
& 
& 
&
& 
& 
& $\frac{E^2}{f_\pi^2}\frac{E^2}{\Lambda^2}\parallel g^2$ 
& $W^-W^\pm \ra W^-W^\pm /(LLLL)~^\ddagger$\\
&
& 
& 
& $\triangle$
& $\triangle$
& 
& 
& $\triangle$
&
& $\triangle$
& $g\frac{E}{f_\pi}\frac{E^2}{\Lambda^2}\parallel g^2\frac{M_W}{E}$ 
& $W^-W^\pm\ra W^-W^\pm /(LLLT)~^\ddagger$ \\
& 
& 
& 
& 
& $\surd$
& $\surd$
&
& 
&
& 
& $\frac{E^2}{f_\pi^2}\frac{E^2}{\Lambda^2}\parallel g^2 $
& $W^-W^+ \ra ZZ ~\&~{\rm perm.}/(LLLL)$ \\
&
& 
& $\triangle$
& $\triangle$
& $\triangle$
& $\triangle$
& 
& $\triangle$
&
& $\triangle$
& $g\frac{E}{f_\pi}\frac{E^2}{\Lambda^2}\parallel g^2\frac{M_W}{E}$ 
& $W^-W^+ \ra ZZ ~\&~{\rm perm.} /(LLLT)$ \\
& 
& 
& 
& 
& $\surd$
& $\surd$
& 
& 
& $\surd$ 
&  
& $\frac{E^2}{f_\pi^2}\frac{E^2}{\Lambda^2}\parallel
   g^2\frac{E^2}{\Lambda^2} $
& $ZZ\ra ZZ /(LLLL) $\\
&
& 
& 
& $\triangle$
& $\triangle$
& $\triangle$
&  
&
& $\triangle$
&
& $g\frac{E}{f_\pi}\frac{E^2}{\Lambda^2}\parallel
  g^2\frac{M_WE}{\Lambda^2}$ 
& $ZZ\ra ZZ /(LLLT) $\\
\hline
& 
& 
& 
& $\surd$
& 
& 
& 
&
& 
& $\surd$
& $g^2\frac{E^2}{\Lambda^2} \parallel  g^2\frac{M^2_W}{E^2}$
& $q\bar{q'}\ra W^\pm Z /(LL) $\\
& 
& $\triangle$
& $\triangle$
& $\triangle$
& 
& 
& $\triangle$
& $\triangle$
&
& $\triangle$
& $g^3\frac{Ef_\pi}{\Lambda^2}\parallel g^2\frac{M_W}{E}$ 
& $q\bar{q'}\ra W^\pm Z /(LT) $\\
 LHC($14$)
& 
& 
& 
& $\surd$
& $\surd$
& 
& 
& $\surd$
&
& $\surd$
& $g^2\frac{1}{f_\pi}\frac{E^2}{\Lambda^2}\parallel g^3\frac{M_W}{E^2}$
& $q \bar{q'}\ra W^-W^+W^\pm /(LLL) $\\
& 
& 
& $\triangle$
& $\triangle$
& $\triangle$
&
& $\triangle$
& $\triangle$
&
& $\triangle$
& $g^3\frac{E}{\Lambda^2}\parallel g^3\frac{M_W^2}{E^3}$ 
& $q \bar{q'} \ra W^- W^+W^\pm  /(LLT)  $\\
& 
& 
& 
& $\surd$
& $\surd$
& $\surd$
& 
& 
&
& $\surd$
& $g^2\frac{1}{f_\pi}\frac{E^2}{\Lambda^2}\parallel g^3\frac{M_W}{E^2}$
& $q \bar{q'}\ra W^\pm ZZ /(LLL) $\\
& 
& $\triangle$
& $\triangle$
& $\triangle$
& $\triangle$
& $\triangle$
& $\triangle$
& $\triangle$
&
& $\triangle$
& $g^3\frac{E}{\Lambda^2}\parallel  g^3\frac{M_W^2}{E^3}$ 
& $q \bar{q'} \ra W^\pm ZZ  /(LLT)  $\\
\hline
 LC($e^-\gamma$)
& 
& $\surd$
& $\surd$
& $\surd$
& 
& 
& $\surd$
& $\surd$
&
& $\surd$
& $eg^2\frac{E}{\Lambda^2}\parallel eg^2\frac{M_W^2}{E^3}$ 
& $e^-\gamma \ra \nu_e W^-Z,e^-WW /(LL)$\\
\hline
& 
& $\surd$
& $\surd$
& $\surd$
& 
& 
& $\surd$
& $\surd$
&
& 
& $e^2\frac{E^2}{\Lambda^2}\parallel e^2\frac{M_W^2}{E^2}$ 
& $\gamma \gamma \ra W^- W^+ /(LL)$\\
LC($\gamma \gamma $)
& 
& $\triangle$
& $\triangle$
& $\triangle$
& 
& 
& $\triangle$
& $\triangle$
&
& 
& $e^2g\frac{Ef_\pi}{\Lambda^2}\parallel e^2\frac{M_W}{E}$  
& $\gamma\gamma \ra W^-W^+ /(LT)$\\
& & & & & & & & & & & & \\
\hline\hline 
\end{tabular}
\clearpage

\newpage
\renewcommand{\baselinestretch}{1.5}
\normalsize
\null\noindent
We see that the coefficients $\ell$'s can be experimentally determined 
via various $V^a_L$ processes at various phases of the LHC and the LC 
(including the $e\gamma$ collider) complementarily. Without the LC, the LHC 
itself is not enough for determining all the coefficients. Quantitative
calculations on the determination of the quartic-$V^a_L$-couplings
$\ell_4$ and $\ell_5$ at the 1.6 TeV LC has been carried out in Ref.\cite{l4l5}.
The results are shown in Fig. 2 which shows that with polarized
electron beams, $\ell_4$ and $\ell_5$ can be determined at a higher accuracy.
Determination of custodial-symmetry-violating-term coefficients $\ell_6$ and 
$\ell_7$ via the interplay between the $V_LV_L$ fusion and $VVV$ production
has been stdied in Ref.\cite{l6l7}.


\centerline{\epsfig{figure=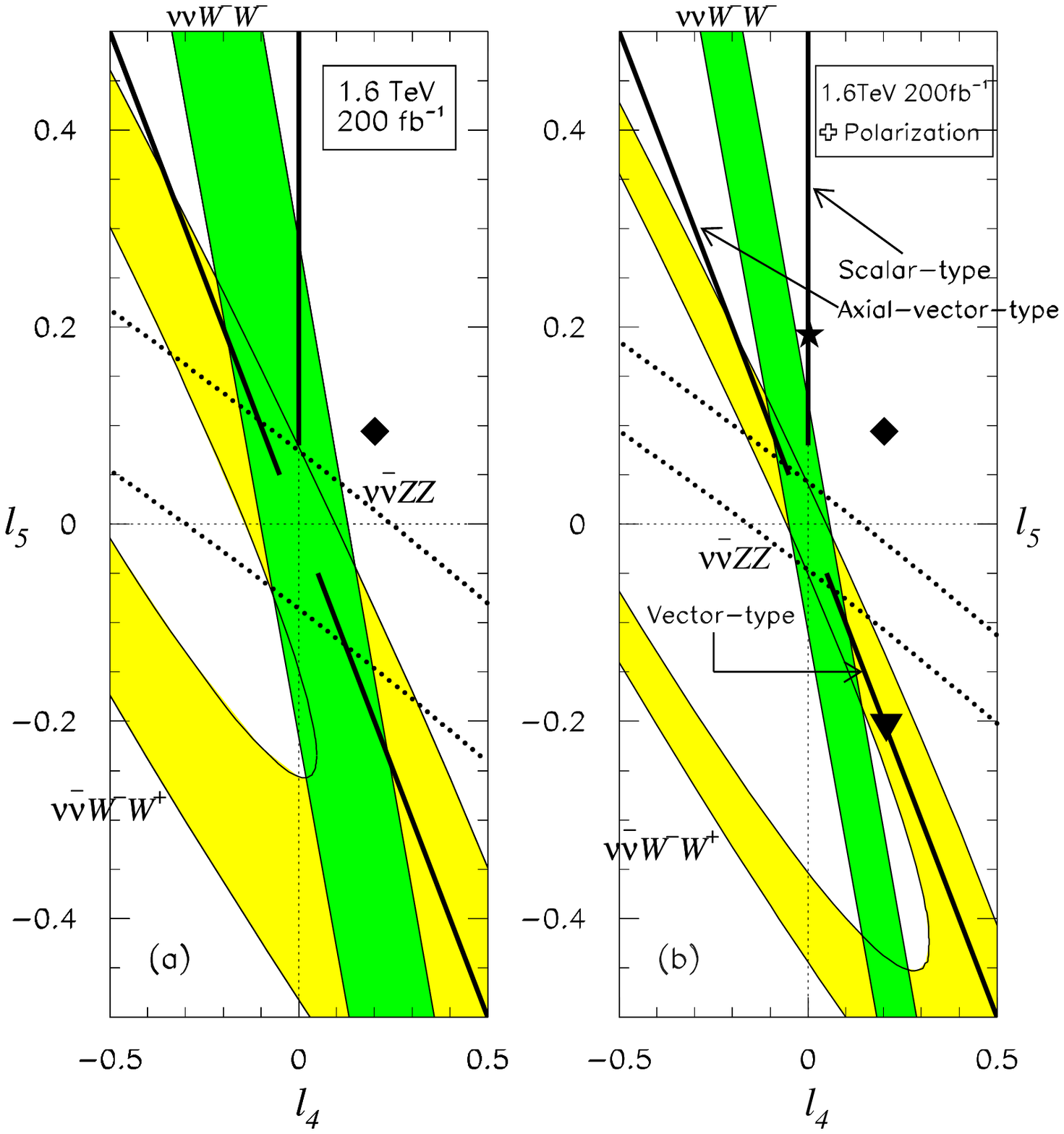,height=8cm}}
\fcaption{Determining the coefficients $\ell_4,~\ell_5$ at the 1.6 TeV
$e^+e^-/e^-e^-$ LC's. The $\pm 1\sigma$ exclusion contours are displayed.
(a) unpolarized case; (b) the case of $90\%(65\%)$ polarized $e^-(e^+)$ beam.
The thick solid lines are contributions from certain simple theoretical
models. Quoted from Ref.\cite{l4l5}.}

\null\vspace{0.2cm}

As we have mentioned, the top quark couples strongly to the EWSB
sector, so that studying effective anomalous couplings of
the top quark will be very helpful for model-independent probe of the
EWSB mechanism. This kind of study has been carried out in 
Refs.\cite{peccei,yuan,young,BPSS}

\null\noindent
\begin{center}
{\bf VI. The Top Quark}
\end{center}

The details of top quark physics are given in the talk by G.P. Yeh at
this workshop. Here we only mention a few topics related to top quark
physics at the LC.

The precise measurements of the properties of the top quark are important for 
testing the SM and probing new physics. For $m_t\sim 175$ GeV, the
maximal cross section of
\begin{eqnarray}
e^+e^-\stackrel{\gamma,Z}{\longrightarrow} t\bar{t}
\label{eett}
\end{eqnarray}
at the LC is
\begin{eqnarray}
\sigma(t\bar{t})\sim 800~{\rm fb}
\end{eqnarray}
at about 30 GeV above the threshold. With an integrated luminosity of
$\int{\cal L}dt=$50 fb$^{-1}$, detailed exoerimental simulations predict
that the to quark mass can be determined to the accuracy of
\begin{eqnarray}
\delta m_t=200~ {\rm MeV}
\end{eqnarray}
at $m_t\sim 180$ GeV \cite{zerwas}. This is better than the claimed
$\delta m_t=3$ GeV at the proton colliders.

If the electron in (\ref{eett}) is left-handed polarized, the
produced top quark is preferentially left-handed in the forward direction
while only a small fraction is produced as right-handed particle in the
backward direction. Thus the backward direction is very sensitive to a
small anomalous magnetic moment of the top quark which can be measured
to a precision of  a few percent \cite{zerwas}. Electric dipole moment
of the top quark can also be well measured \cite{zerwas}.

Recent Fermilab CDF data on branching ratio of $~t\to b+W^+\to b+l^++\nu_l~$ 
is
\begin{eqnarray}
B(t\to bl^+\nu_l)=0.188\pm 0.048\,,
\end{eqnarray}
which is consistent with the tree-level SM prediction 2/9
\cite{tdecay}. Further precise measurements can give constraints on new
physics. For instance, it can give constraint on $\tan\beta$ in the
MSSM since $t\to b+H^+$ and $t\to \tilde{b}+\tilde{\chi}^+_1$ increase rapidly 
with $\tan\beta$ \cite{sola}; it can give constraint on the technipion
and top-pion masses since $~t\to \pi^+,\pi^+_t~$ will be large for light 
$\pi^+$ and $\pi_t^+$.

It has been expected that the threshold measurement of $t\bar{t}$
production at the LC can provide precision measurements of $m_t$ and 
$\alpha_s$ due to the large cross section near the threshold. However, 
there are subtleties in the theoretical calculation.
Near the threshold, both $\alpha_s$ and the relative velocity $v$ are
small, and $\alpha_s/v$ is of the order of 1 and have to be summed up
to all orders even in the leading order approximation. To 
next-to-leading order (NLO) and next-to-next-to-leading order (NNLO), one has 
to resum all $(\alpha_s/v)^N(1,\alpha_s,v,\alpha_s^2,\alpha_sv,v^2)$
terms. Recent calculations of the resummation \cite{yakovlev} show that the 
NNLO terms are of the same order as the NLO terms ($\sim 10\%$) (cf. Fig. 3). 
So that the convergence is actually not good, and thus the theoretical 
uncertainty cannot be so small as expected.

\null\vspace{-0.6cm}
\centerline{\epsfig{figure=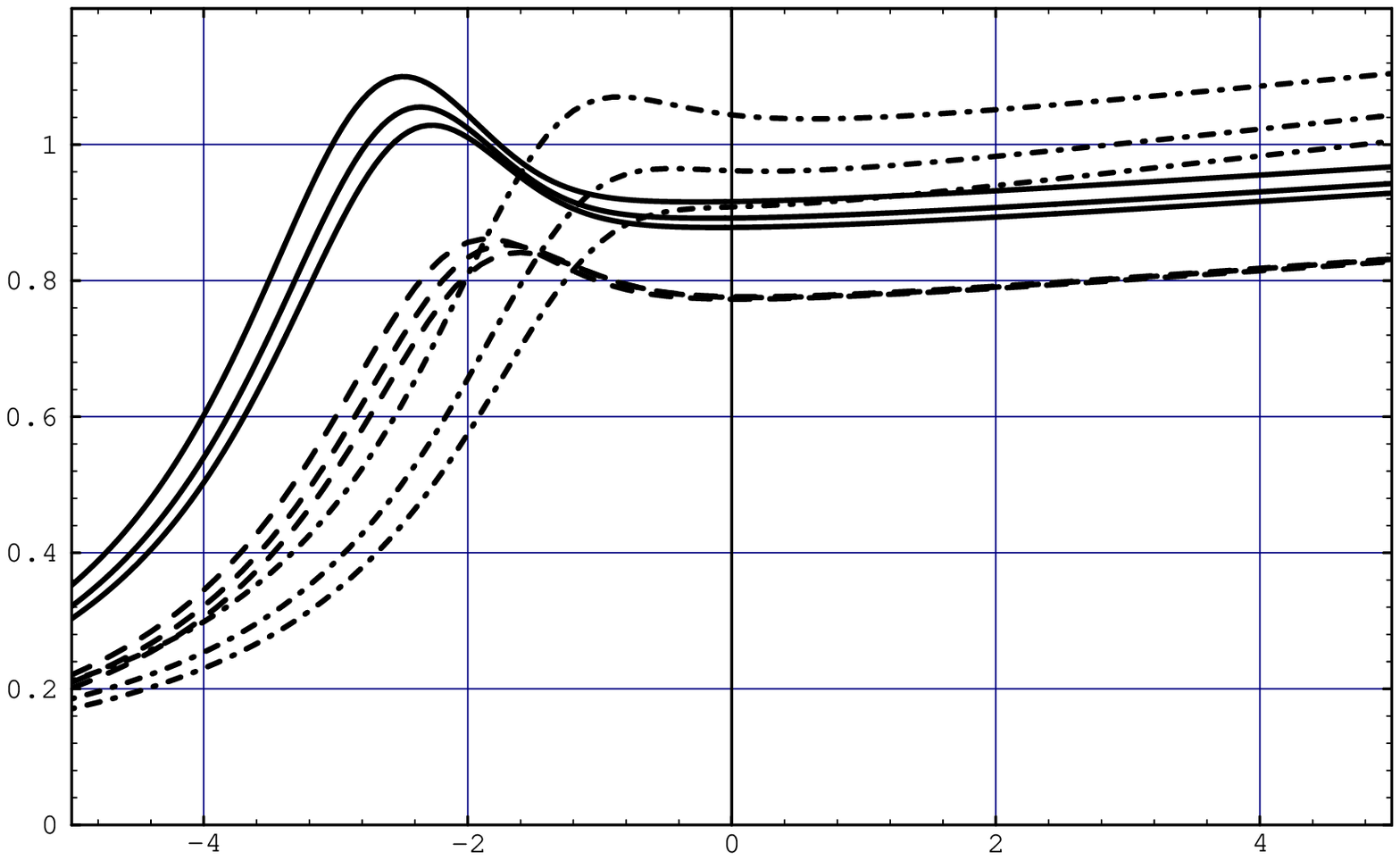,height=6cm}}
\vspace{-0.6cm}
\fcaption{The LO (dashed-dotted lines), NLO (dashed lines) and
NNLO (solid lines) contributions to the rate of $e^+e^-\to t\bar{t}$.
Quoted from Ref.\cite{yakovlev}.}
\null\vspace{0.4cm}

In view of the large $t\bar{t}$ production cross section at the photoh-photon 
collider, another recent investigation suggested to do the precision 
measurements via $\gamma\gamma\to t\bar{t}$ off the threshold region to
avoid the difficult threshold calculation \cite{kamal}. The calculation
in Ref.\cite{kamal} shows that the optimal energy is 
$\sqrt{s_{\gamma\gamma}}=420 $ GeV which should be accessible at a 
$\sqrt{s_{e^+e^-}}\agt 500$ GeV LC, and at this collider with a typical 
$\gamma\gamma$ integrated luminosity of 20 fb$^{-1}$,
$\alpha_s$ can be determined at the accuracy of $3\%$ statistically 
\cite{kamal}. A better accuracy can be achieved with a larger luminosity.

\vspace{1.2cm}
\begin{center}
{\bf VII. Conclusions}
\end{center}

The SM has succssfully passed a lot of precision experimental tests, but 
there are still certain notable deviation when doing the
model-independent analyses, namely the $3\sigma$ deviation of the right-handed
effective coupling of the $b$-quark. At present, we cannot draw 
conclusion on the need of new physics beyond the SM only from this deviation.
Clearer tests of new physics will be done at future high energy collider 
experiments.

Despite of the success of the SM, its EWSB sector is still not clear. The 
assumed Higgs boson in the SM has not been found, and the SM Higgs sector 
suffers from the well-known problems of {\it triviality} and 
{\it unnaturalness}, so that the EWSB sector may concern new physics. Although 
the global fit of the SM to the precision electroweak data favors a light 
Higgs boson, there are possible new physics models without a light Higgs boson 
that can also fit the precision data. So that the search for the Higgs boson 
should be carried out in the whole possible energy range up to 1 TeV. 

Since the EWSB mechanism concerns the understanding of the origin of 
particle masses, the probe of it is a very interesting and importnat topic 
in current particle physics. If a light Higgs boson (elementary or composite)
exists, it can be found, as we have seen, at the future high energy colliders 
such as the LHC, the LC (including the $\gamma\gamma$ and $e\gamma$ colliders),
etc. The LC has the advantage of low hadronic backgrounds. After
finding the Higgs boson, we have to further study its properties to see
if it is just the SM Higgs boson, or a Higgs boson in a more
complicated new physics model (e.g. the MSSM), or it is composite. If
there is no light Higgs boson, a feasible way of probing the EWSB
mechanism is to study the longitudinal weak boson reactions and
$t\bar{t}$ productions. We have seen that, for this purpose, the LHC
alone is not enough and the LC (including the $\gamma\gamma$ and
$e\gamma$ colliders) are needed.

SUSY and strongly interacting EWSB mechanism are two possible
candidates of new physics which can avoid the shortcomings of the SM Higgs 
sector. The most convincing way of testing SUSY is to search for SUSY 
particles at high energy colliders. If a SUSY particle is found, we should 
study its properties to see if the SUSY model is just the simplest MSSM or 
more complicated ones. An interesting conclusion is that if a light Higgs 
boson is not found below 130 GeV, the MSSM will be in a bad shape and
non-minimal SUSY models will be favored. After finding the SUSY particles, our 
direction may be lead to SUSY GUT and superstring to find out the origin of 
the large number of free parameters in the SUSY model. If no SUSY particles
are found below 1 TeV, SUSY may not be relevant to the solution of
the unnaturalness problem. Then the possible solution may be the
strongly interacting EWSB mechanism in which both the unnaturalness and
triviality problems no long exist. In this case, our direction may be lead to
the study of new strong interactions and new composite particles above
1 TeV. In all these studies, the LC will play an important role.

Because of its large mass, the top quark couples strongly to the EWSM
sector, so that it plays an important role in probing the EWSB mechanism.
In addition, high precision study of the properties of the top quark
is important in testing the SM and studying new physics. This can be
effectively done at the LC (including the $\gamma\gamma$ and $e\gamma$ 
colliders).

In summary, particle physics will be in a crucial status of clarifying
the choice of different directions of new physics when we go to the TeV energy 
scale. The LC will be an important equipment for studying TeV physics and 
will help us to know to which direction we should further go. Further
theoretical and experimental studies of LC physics are needed.

\vspace{0.5cm}
\null\noindent
{\bf Acknowledgement}

This work is supported by the National Natural Science Foundation of China,
The foundation of Fundamental Research of Tsinghua University, and a
special grant from the Chinese Ministry of Education.


\vspace{1.2cm}
\noindent
\renewcommand{\baselinestretch}{1.2}

\end{document}